\documentclass{iau}
\usepackage{graphicx}

\title[Comparison between maser and optical late-type stars] 
{Maser, infrared and optical emission for late-type stars in the Galactic plane}
\author[Quiroga-Nu\~{n}ez et al.]   
{L.H.~Quiroga-Nu\~{n}ez$^{1,2}$,
    H.J.~van~Langevelde$^{2,1}$,   
	L.O.~Sjouwerman$^{3}$,
	Y.M.~Pihlstr\"{o}m$^{4}$,
	M.J.~Reid$^{5}$,    
	A.G.A.~Brown$^{1}$    
 	\and J.A.~Green$^{6}$
}

\affiliation{$^1$ Leiden Observatory, Leiden University, P.O. Box 9513, Leiden, The Netherlands. \\ email: {\tt quiroganunez@strw.leidenuniv.nl} \\[\affilskip]
$^2$Joint Institute for VLBI ERIC (JIVE), Postbus 2, Dwingeloo, The Netherlands. \\[\affilskip]
$^3$National Radio Astronomy Observatory, P.O. Box 0, Lopezville Road 1001, Socorro, USA. \\ [\affilskip]
$^4$Department of Physics and Astronomy, University of New Mexico, MSC07 4220, Albuquerque, USA. \\ [\affilskip]
$^5$Harvard-Smithsonian Center for Astrophysics, 60 Garden Street, Cambridge, USA. \\ [\affilskip]
$^6$CSIRO Astronomy and Space Science, Australia Telescope National Facility, P.O. Box 76, Epping, Australia. \\ [\affilskip]
}

\pubyear{2017}
\volume{336}  
\jname{Astrophysical Masers: Unlocking the Mysteries of the Universe}
\editors{A. Tarchi, M.J. Reid \& P. Castangia}
\begin{document}

\maketitle
\vspace*{-0.2 cm}
\begin{abstract}
Radio astrometric campaigns using VLBI have provided distances and proper motions for masers associated with young massive stars (BeSSeL survey). The ongoing BAaDE project plans to obtain astrometric information of SiO maser stars located in the inner Galaxy. These stars are associated with evolved, mass-losing stars. By overlapping optical (\textit{Gaia}), infrared (2MASS, MSX and WISE) and radio (BAaDE) sources, we expect to obtain important clues on the intrinsic properties and population distribution of late-type stars. Moreover, a comparison of the Galactic parameters obtained with \textit{Gaia} and VLBI can be done using radio observations on different targets: young massive stars (BeSSeL) and evolved stars (BAaDE).
\end{abstract}

\vspace*{-0.3 cm}
\firstsection 
\section{Context: Maser surveys and \textit{Gaia} counterparts}

Molecular masers have been used for decades to study astrophysical environments~(see e.g.,~\cite{Elitzur92,Elitzur05} and references therein). Since then, radio campaigns have used masers in order to provide insights into the structure of the Milky Way. In this study, we focus on two specific radio surveys that aim to study different regions of the Galaxy: the spiral structure and the Galactic bulge.

By using data from the Very Large Baseline Array (VLBA) and the European VLBI Network (EVN), the Bar and Spiral Structure Legacy (BeSSeL) survey has obtained accurate astrometric measurements with a resolution up to $\sim$10$\mu$as. The BeSSeL survey targeted high-mass star forming regions (HMSFRs) uniquely associated with water (22 GHz) and methanol maser (6.7 and 12 GHz) emission. By fitting 6D phase-space information of the observed HMSFRs to a spiral Galactic model, BeSSeL has determined the fundamental Galactic parameters and the solar motion with respect to the Local Standard of Rest~(\cite{Reid14b}). Additionally, simulated data of several Galactic distributions of methanol maser bearing stars has confirmed the accuracy of the Galactic parameter values found by the BeSSeL survey~(\cite{Quiroga17}).

The Bulge Asymmetries and Dynamical Evolution (BAaDE) project is a survey for SiO maser bearing stars in the Galactic plane, focusing on the bulge~(\cite{Sjouwerman17}). By selecting targets based on their infrared color using MSX data~(\cite{Sjouwerman09}), more than 20,000 evolved stars have been searched for SiO maser lines using the VLA (43 GHz) and ALMA (86 GHz). Thousands of line-of-sight velocities and peak flux densities, along with other line properties, for late-type stars in the inner Galaxy are being compiled.

\begin{figure}[t!]
\vspace*{-0.25 cm}
\resizebox{\hsize}{!}{\includegraphics{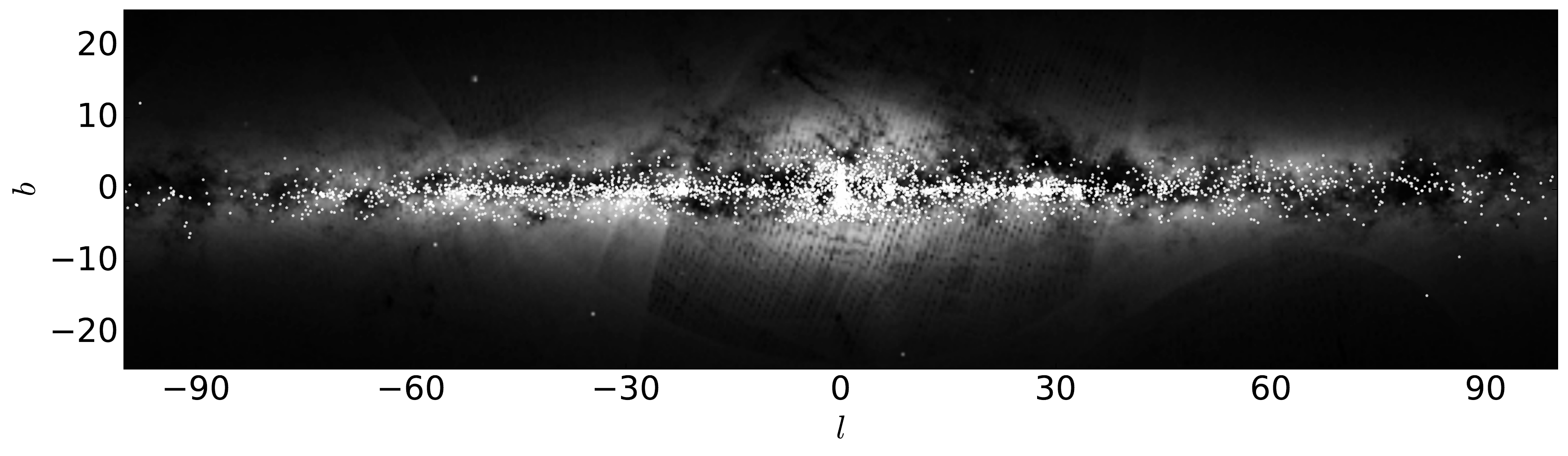}}
\vspace*{-0.6 cm}
\caption{Galactic distribution of the sample obtained by cross-matching BAaDE targets, 2MASS and \textit{Gaia} DR1 over the Gaia first sky map. Credit: ESA/Gaia/DPAC.}
\vspace*{-0.15cm}
\label{fig1}
\end{figure}

Although BAaDE is mapping Galactic regions where optical surveys typically do not reach ($|b|\leq 5^{o}$), a refined positional cross-match between \textit{Gaia} DR1, 2MASS and BAaDE sources has shown more than 2,000 coincidences (see Fig.~\ref{fig1}), which represents~$\sim$8$\%$ of the BAaDE sample~($\sim$30,000 targets).
Moreover, around 150 sources of the cross-matched sample are part of the Tycho catalog, and therefore, an estimate of their parallaxes are already available in Tycho-Gaia Astrometric Solution (TGAS).

\vspace*{-0.4cm}

\section{Future studies with the cross-matched sample}
1. \textit{Gaia} DR2 will soon provide optical photometry, parallaxes, proper motions and periods that will complement the radio (BAaDE) and infrared (2MASS, MSX and WISE) data known for these 2,000 late-type stars. Using these data, we aim to characterize the stellar population in the Galactic bulge by providing statistics of stellar masses, ages, metallicities, periods and luminosities. It is thought that the stellar population at the bulge can provide clues to understand the dynamical evolution of the Milky Way.

2. VLBI astrometry of some bright BAaDE targets is being defined to provide parallaxes and 3D orbits with an estimated accuracy of~$\sim$50$\mu$as. By doing VLBI, we can refine a subset of the orbits and perhaps get the orbit class, and therefore, signatures of past mergers of other stellar systems with the Galaxy.

3. Astrometric VLBI results would also allow a direct comparison of the parallax technique between optical (\textit{Gaia} DR2) and radio data, which will align optical stellar images with SiO maser rings for nearby sources. An initial comparison is ongoing using the parallaxes from TGAS.

4. By getting full phase-space information of a subset of evolved stars in the Galactic bulge, dynamical models for the inner Galaxy can be tested. Orbital motions affected by the bar can be fitted to refine the fundamental parameter values of the Milky Way.
\vspace*{-0.15cm}
\begin{acknowledgements} 
This work has made use of data from the ESA mission \textit{Gaia}, processed by DPAC. Funding for DPAC has been provided by national institutions participating in the \textit{Gaia} Multilateral Agreement.
\end{acknowledgements} 


\vspace*{-0.5cm}
\begin{thebibliography}{}

\bibitem[Elitzur, 1992]{Elitzur92}
{Elitzur, M.} 1992, \textit{ARA$\&$A}, 30, 75

\bibitem[Elitzur, 2005]{Elitzur05}
{Elitzur, M.} 2005, \textit{Science}, 309, 71

\bibitem[Quiroga-Nu\~nez et al., 2017]{Quiroga17}
{Quiroga-Nu\~nez, L.~H., van Langevelde, H.~J., Reid, M.~J. \& Green, J.~A.} 2017, \textit{A$\&$A}, 604, A72

\bibitem[Reid et al., 2014]{Reid14b}
{Reid, M.~J., Menten, K.~M., Brunthaler, Zheng, X.~W., et al.} 2014, \textit{ApJ}, 783, 130

\bibitem[Sjouwerman et al., 2009]{Sjouwerman09}
{Sjouwerman, L.~O., Capen, S.~M. $\&$ Claussen, M.~J.} 2009, \textit{ApJ}, 705, 1554

\bibitem[Sjouwerman et al., 2017]{Sjouwerman17}
{Sjouwerman, L.~O., Pihlstr\"{o}m, Y.~M., Rich, R.~M., et al.} 2016, \textit{Proc. IAU Symposium}, 322, 103 

\end{thebibliography}
\end{document}